\documentclass[final,5p]{elsarticle}
\usepackage{graphicx}
\usepackage{amssymb,amsmath}
\usepackage[colorlinks=true]{hyperref}
\biboptions{comma,square}

\newcommand{\rmd}{\mathrm{d}}

\journal{Physica Medica}
\begin{document}

\begin{frontmatter}
\title{Dose calculation algorithm of fast fine-heterogeneity correction for heavy charged particle radiotherapy}
\author[ad1,ad2]{Nobuyuki Kanematsu}
\ead{nkanemat@nirs.go.jp}
\address[ad1]{Department of Accelerator and Medical Physics, Research Center for Charged Particle Therapy, National Institute of Radiological Sciences, 4-9-1 Anagawa, Inage-ku, Chiba 263-8555, Japan}
\address[ad2]{Department of Quantum Science and Energy Engineering, Tohoku University, 6-6 Aramaki Aza Aoba, Aoba-ku, Sendai 980-8579, Japan}
\begin{abstract}
This work addresses computing techniques for dose calculations in treatment planning with proton and ion beams, based on an efficient kernel-convolution method referred to as grid-dose spreading (GDS) and accurate heterogeneity-correction method referred to as Gaussian beam splitting.
The original GDS algorithm suffered from distortion of dose distribution for beams tilted with respect to the dose-grid axes.
Use of intermediate grids normal to the beam field has solved the beam-tilting distortion.
Interplay of arrangement between beams and grids was found as another intrinsic source of artifact. 
Inclusion of rectangular-kernel convolution in beam transport, to share the beam contribution among the nearest grids in a regulatory manner, has solved the interplay problem.
This algorithmic framework was applied to a tilted proton pencil beam and a broad carbon-ion beam.
In these cases, while the elementary pencil beams individually split into several tens, the calculation time increased only by several times with the GDS algorithm.
The GDS and beam-splitting methods will complementarily enable accurate and efficient dose calculations for radiotherapy with protons and ions.
\end{abstract}
\end{frontmatter}

\section{Introduction}

Dose distributions of radiotherapy are represented by point doses at orthogonally arranged grids. 
In treatment-planning practice, the grid intervals are defined from a physical, clinical, and practical points of view, often resulting in cubic dimensions of a few millimeters.
Accuracy, efficiency and their balance are essential in practice, for which the pencil-beam algorithm is commonly used. That is mathematically a convolution integral of total energy released per mass (terma) with elementary beam-spread kernel, which may be computationally demanding.

The grid-dose-spreading (GDS) algorithm was developed for fast dose calculation of heavy-charged-particle beams in patient body \cite{Kanematsu2008}.
The GDS algorithm employs approximation to extract beam-interaction part from the integral at the expense of distortion of dose distribution for a beam tilted with respect to the grid axes, as originally recognized in Ref.~\cite{Kanematsu2008}.
The beam-tilting distortion may be generally insignificant when beam blurring is as small as the required spatial resolution, for example, for a carbon-ion beam.
In fact, the GDS method was successfully incorporated into a clinical treatment-planning system for carbon-ion radiotherapy with vertical and horizontal fixed beams \cite{Endo1996, Kanematsu2002}, for which tilting was intrinsically absent.

In that particular implementation, a simplistic post process was added to the original broad-beam algorithm so as to spread an intermediate terma distribution uniformly \cite{Kanematsu2008}.
In general, the spreading kernel could be spatially modulated using the pencil-beam model for more accurate heterogeneity correction \cite{Petti1992}.
There are two reciprocal approaches for convolution, {\it i.e.} to collect doses transferred from nearby interactions to a grid or \emph{the dose-deposition point of view} and to spread a terma from an interaction to nearby grids or \emph{the interaction point of view}.
The latter is usually more efficient than the former for three-dimensional dose calculation \cite{Mackie1985}.

\begin{figure}
\includegraphics[width=8.5cm]{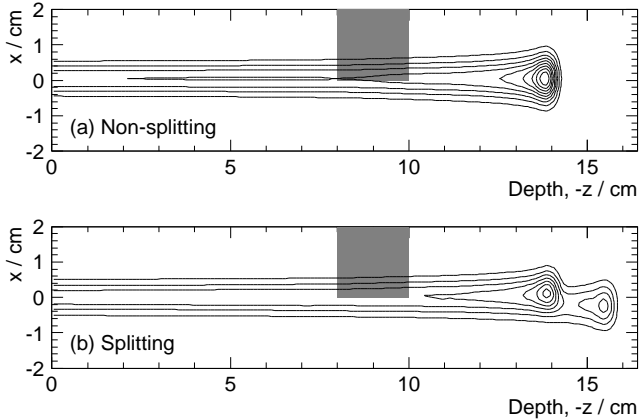}
\caption{\label{fig:split} (a) Non-splitting and (b) splitting dose calculations with isodose lines at every 10\% levels of the maximum non-splitting dose in the $y = 0$ cross section, where a proton pencil beam with $E = 150$ MeV and $\sigma = 3$ mm is incident into water with a bone-equivalent material ($\rho = 1.8~{\rm g}/{\rm cm}^3$) inserted halfway (gray area).}
\end{figure}

The pencil-beam model implicitly assumes homogeneity of the medium within the elementary beam spread. 
Beams that have grown excessively thick in heterogeneous transport are thus incompatible. 
As a general and rigorous solution, Gaussian-beam splitting was proposed, with which overgrown beams are subdivided into smaller ones at locations of large lateral heterogeneity \cite{Kanematsu2009b}.
Figure \ref{fig:split} demonstrates its effectiveness for a simple density boundary, where the non-splitting beam happened to traverse an edge of a bone-equivalent material while about a half of the split beams traverse the bone-equivalent material.
The splitting causes explosive beam multiplication in a shower-like process.
In this particular case for example, the original beam recursively split into 28 final beams.
Slowing down of dose calculation due to beam multiplication will be a problem in practice.

In Ref.~\cite{Kanematsu2009b}, the beam-splitting method was stated as efficient due to certain ``algorithmic techniques to be explained elsewhere'', which in fact implied this work to construct a framework, where the GDS and beam-splitting methods work compatibly for accurate and efficient dose calculations.
In addition, we will refine the GDS algorithm with a fix against the beam-tilting distortion and with the pencil-beam model in the interaction point of view for better heterogeneity correction.

Although the Gaussian-beam approximation may be reasonable for the multiple-scattering effect, two or more Gaussian components would improve the accuracy of lateral dose distribution of proton and ion pencil beams \cite{Pedroni2005, Inaniwa2009}.
However, such large-sized components are intrinsically incompatible with fine heterogeneity. 
In addition, it is inconceivable to apply the beam-splitting method for large-sized components to secure practical efficiency.

This framework will be applicable not only to broad-beam delivery but also to pencil-beam scanning, where a physical scanned beam may have to be decomposed into virtual elementary beams to address heterogeneity \cite{Schaffner1999}.
As this work aims to improve computing methods, we focus on evaluation of efficiency and settlement of the intrinsic artifacts with respect to the ideal beam models that are mathematically given, without repeating experimental assessments of accuracy \cite{Kanematsu2009b}.

\section{Materials and methods}

\subsection{Algorithmic techniques}

\subsubsection{Grid normalization}

\begin{figure}\begin{center}
\includegraphics{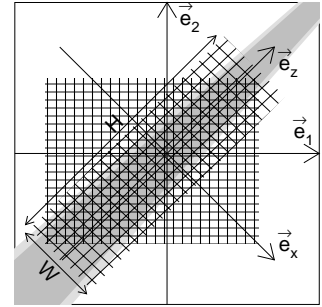}\end{center}
\caption{\label{fig:coordinates}Schematic of grid normalization in cross-section view perpendicular to $\vec{e}_3 = \vec{e}_y$, where shown are a field (gray area) with axes $\vec{e}_x$ and $\vec{e}_z$, margins (light gray areas), original grids with axes $\vec{e}_1$ and $\vec{e}_2$, and normal grids of width $W$ and height $H$.}
\end{figure}

We will solve the beam-tilting distortion of the GDS algorithm by defining intermediate grids for dose calculation, which are arranged to be normal to the beam-field axes.
As shown in Figure \ref{fig:coordinates}, the original dose grids along numbered axes 1, 2, and 3 are defined with basis vectors $\vec{e}_1$, $\vec{e}_2$, and $\vec{e}_3$ and intervals $\delta_1$, $\delta_2$, and $\delta_3$.
For a given radiation field, the field coordinates $x$, $y$, and $z$ with basis vectors $\vec{e}_x$, $\vec{e}_y$, and $\vec{e}_z$ are associated, where the origin is at the isocenter and $\vec{e}_z$ is in the source direction.
With lateral margins for penumbra, the normal-grid volume is defined as the supremum of normal rectangular-parallelepiped volume of $W\times L\times H$ containing the original grids in the margined field.
Quadratic projection of the original-grid voxel gives the normal-grid intervals $\delta_x$, $\delta_y$, and $\delta_z$ as
\begin{equation}
\delta_{\{x,y,z\}}^2 = \sum_{a=1}^3 \delta_a^2 \left(\vec{e}_a \cdot \vec{e}_{\{x,y,z\}}\right)^2,
\end{equation}
to approximately conserve the equivalent resolution.
Normal grids $\hat{g}_{i j k}$ are defined at equally spaced positions $\vec{r}_{\hat{g}_{i j k}}$ for indices $i \in [0, \lceil W/\delta_x \rceil]$, $j \in [0, \lceil L/\delta_y \rceil]$ and $k \in [0, \lceil H/\delta_z \rceil]$, where $\lceil\ \rceil$ is the ceiling function.

Because the elementary pencil beams are almost parallel to vector $\vec{e}_z$, normal-grid doses $D_{\hat{g}}$ can be accurately and efficiently calculated with
\begin{equation}
D_{\hat{g}} = \sum_{\hat{h}} T_{\hat{h}}
\int_{x_{\hat{g}}-\delta_x/2}^{x_{\hat{g}}+\delta_x/2} G_{x_{\hat{h}},\sigma_{\hat{h}}}(x)\, \rmd x
\int_{y_{\hat{g}}-\delta_y/2}^{y_{\hat{g}}+\delta_y/2} G_{y_{\hat{h}},\sigma_{\hat{h}}}(y)\, \rmd y,
\label{eq:spreading}
\end{equation}
where $T_{\hat{h}}$, $\sigma_{\hat{h}}$, and $(x_{\hat{h}},y_{\hat{h}})$ are the terma, the spread, and the position of normal grid $\hat{h}$, and $G_{m,\sigma}(x)$ is the Gaussian distribution of mean $m$ and standard deviation $\sigma$ whose integral is readily given by the standard error function.
Dose $D_g$ at original grid $g$ is then given by trilinear interpolation,
which is repeated for all the original grids in the margined field.
In this manner, we only deal with normal grids in the GDS algorithm hereafter.

\subsubsection{Inter-grid beam sharing}

\begin{figure}
\includegraphics[width=8.5cm]{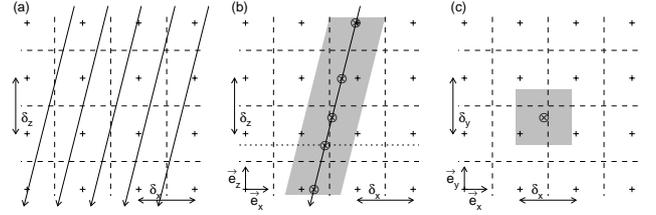}
\caption{\label{fig:sharing}Schematic of inter-grid beam sharing in the cross-section views with grid points ({\tiny +}), voxel boundaries (dashed lines), beam paths (arrows), mid-step points ($\otimes$), and rectangular $\mathbf{A}_b$ (gray areas). One of many beams in (a) is focused in (b), where the dotted line indicates the $x$--$y$ cross section in (c).}
\end{figure}

Beam-$b$ spread $\sigma_{b,g}$ at grid $g$ is given by beam-transport calculation \cite{Eyges1948, Kanematsu2009a} and terma contribution $\Delta T_{b,g}$ is defined \cite{Kanematsu2009b} as 
\begin{equation}
\Delta T_{b,g} = \frac{n_b}{\delta_x \delta_y \delta_z} \tilde{D}_{\Phi 0} \, \Delta s,
\label{eq:terma}
\end{equation}
where $n_b$ is the number of particles that is modeled as a constant, $\tilde{D}_{\Phi 0}$ is the dose per in-air fluence that is measured as a depth--dose curve, and $\Delta s$ is the step length in voxel $g$.
In the original GDS algorithm, grid terma $T_g$ and spread $\sigma_g$ are intermediately defined with formulas
\begin{equation}
T_g = \sum_{b} \Delta T_{b,g}, \qquad 
\sigma_g^2 = \frac{1}{T_g}\, \sum_{b} \Delta T_{b,g}\, \sigma_{b,g}^2,
\label{eq:original}
\end{equation}
which are straightforward extensions of Eqs.~(13) and (14) in Ref.~\cite{Kanematsu2008}.
That would be, however, problematic when the beams are arranged independently to the grids.
As shown in Figure \ref{fig:sharing}(a), only several beams may traverse each voxel. Due to interplay between the beam and grid structures, the total beam-path length per voxel and consequently the grid terma $T_g$ largely fluctuates.
The fluctuation would be smeared out only with large spreading with $\sigma_g \gg \delta_{\{x,y\}}$.

To fully resolve the fluctuation, we will distribute step terma $\Delta T_{b,g}$ to the nearest four grids in a regulatory manner.
As shown in Figures \ref{fig:sharing}(b) and \ref{fig:sharing}(c), the $x$ and $y$ axes comprise the lateral plane, where any beam $b$ is now modeled to have cross section $\mathbf{A}_b$ of size $\delta_x \times \delta_y$ centered at mid-step point $\vec{r}_b = (x_b, y_b, z_b)$ in current voxel $\mathbf{V}_h$ at grid $h$.
The four voxels that intersect $\mathbf{A}_b$ share terma contribution $\Delta T_{b,h}$ by areal fractions.
Equation (\ref{eq:original}) is then modified in such a way that the terma and spread distributions will be formed when all the beams have been processed as
\begin{eqnarray}
T_g &=& \sum_b \sum_h \Delta T_{b,h} \, \frac{\mathbf{A}_b \cap \mathbf{V}_g}{\mathbf{A}_b},
\label{eq:tg}
\\
\sigma_g^2 &=& \frac{1}{T_g} \sum_b \sum_h \Delta T_{b,h} \, \sigma_{b,h}^2 \, \frac{\mathbf{A}_b \cap \mathbf{V}_g}{\mathbf{A}_b},
\label{eq:sigmag}
\end{eqnarray}
where the areal fraction is given by
\begin{equation}
\frac{\mathbf{A}_b \cap \mathbf{V}_g}{\mathbf{A}_b} = \left(1-\left|\frac{x_b-x_g}{\delta_x}\right|\right)\cdot\left(1-\left|\frac{y_b-y_g}{\delta_y}\right|\right)
\end{equation}
for nearby grid $g$ with $|x_b-x_g| < \delta_x$ and $|y_b-y_g| < \delta_y$ or otherwise 0. 

Essentially, this beam-sharing operation is one form of convolution with a rectangular kernel.
We thus modify axial dose spreading $\sigma_{h}$ at grid $h$ in (\ref{eq:spreading}) to correct the extra rectangular spreading as
\begin{eqnarray}
\sigma_h \to \sqrt{\sigma_h^2-\delta_{\{x,y\}}^2/12},
\end{eqnarray}
for which, sufficiently fine grid intervals $\delta_{\{x,y\}}$ must be given.

\subsection{Evaluation}

\paragraph{Tilted pencil beam} 

We examine the effect of grid normalization using an analytic 150 MeV proton pencil beam model with formulated dose per in-air fluence \cite{Bortfeld1997} and lateral spread \cite{Kanematsu2009a}.
The beam with rms projected angle $\theta_0 = 10$ mrad and size $\sigma_0 = 1.2$ mm was placed at the isocenter to incident into water in the direction of patient-support angle $\theta_{\rm s} = 0^\circ$ and gantry angle $\phi_{\rm g} = 30^\circ$ in the standard coordinate system \cite{IEC61217}.
The original grids were defined cubically as $\delta_1 = \delta_2 = \delta_3 = 1$ mm to form a volume of 10 cm $\times$ 15 cm $\times$ 10 cm.
Around the beam axis, 1-cm margins were added to define the normal-grid volume.
The dose distributions were calculated using the GDS algorithm with and without grid normalization for comparison. 

\paragraph{Broad beam with heterogeneity}

We examine a broad beam in a heterogeneous medium, which approximates clinical situations.
The dose per in-air fluence was modeled based on an experiment \cite{Kanematsu2009b}, where a carbon-ion beam with incident nucleon kinetic energy $E/A = 290$ MeV was broadened by spiral wobbling \cite{Yonai2008} and was range-modulated by a semi-Gaussian filter of $\sigma_R = 1.8$ mm.
In the calculation, a radiation from a source at height 500 cm was limited by a collimator at height 50 cm to form a 4 cm $\times$ 4 cm field on the isocenter plane, which yielded $81\times81$ original pencil beams.
In a 20 cm $\times$ 20 cm $\times$ 10 cm volume gridded at intervals of 1 mm along axes 1, 2, and 3, an 19 cm diameter cylindrical water phantom was defined at the center.
The phantom included a 2 cm diameter dense ($\rho = 2~{\rm g}/{\rm cm}^3$) rod at radius 6 cm.
The dosimetric effects of grid normalization and inter-grid sharing were investigated for two gantry angles $\phi_{\rm g} = 0^\circ$ and $45^\circ$ with the concurrently rotated phantom so that the rod was always in the middle of the field to see any angular artifact.
The computing times were measured using 2.4 GHz Intel Core2Duo processor on Apple MacBook computer.

%

\section{Results}

\paragraph{Tilted pencil beam}

\begin{figure}
\includegraphics[width=8.5cm]{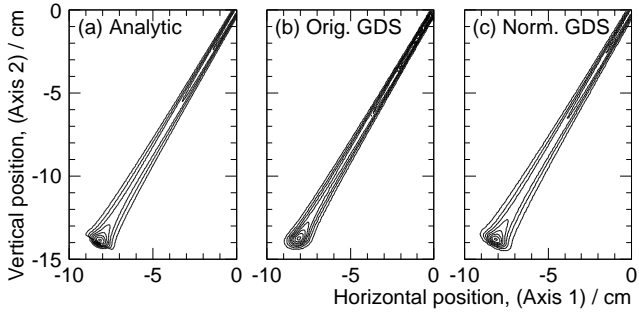}
\caption{\label{fig:pencil}Projected dose at every 10\% levels for (a) the analytic proton-beam model, (b) the original GDS calculation, and (c) the normalized GDS calculation.}
\end{figure}

Figure \ref{fig:pencil} shows the projected dose distributions for the tilted proton pencil beam.
The original GDS algorithm severely distorted the analytic model that was exactly the dose kernel of the algorithm.
The grid normalization greatly reduced the distortion.

\paragraph{Heterogeneous broad beam}

\begin{figure}
\includegraphics[width=8.5cm]{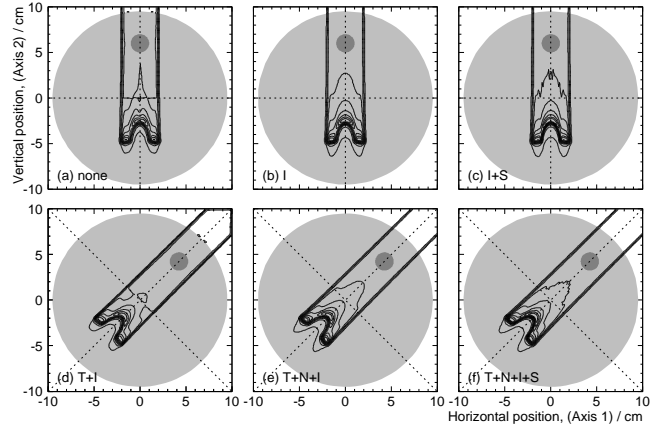}
\caption{\label{fig:broad} Dose distributions in the $y=0$ cross section for a carbon-ion beam calculated in configurations A--F.
The solid and dotted lines indicate every 10\% dose levels and the field $x$ and $z$ axes.
The light-gray and medium-gray areas indicate the water phantom and the high-density rod.}
\end{figure}

\begin{figure}
\includegraphics[width=8.5cm]{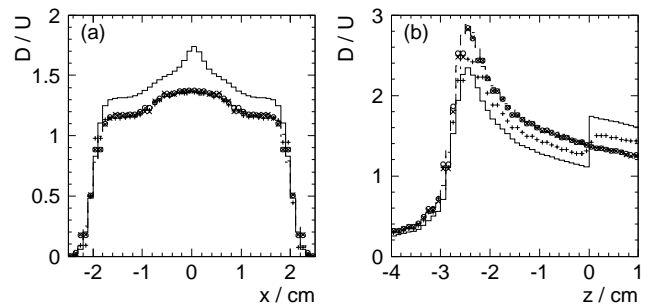}
\caption{\label{fig:broadp} Dose profiles along the (a) $x$ and (b) $z$ axes for a carbon-ion beam in configurations A (solid), B (dashed), C (dotted), D ($+$), E ($\circ$), and F ($\times$) in units of arbitrary reference dose $U$.}
\end{figure}

Table \ref{tab:configurations} and Figures \ref{fig:broad} and \ref{fig:broadp} show the configurations for broad carbon-ion beam calculations, the computing times, and the dose distribution in the $y=0$ cross-section and along the $x$ and $z$ axes of the beam field.
Configurations A and D resulted in severe dose artifacts for the beam--grid interplay and for the tilted incidence while the others are hardly distinguishable. 
From the A--B comparison, we find that the inter-grid sharing resolved the interplay artifact and added marginal computing load. 
From the D--E comparison, we find that the grid normalization resolved the angle issue and unexpectedly reduced the computing time.
From the B--C and E--F comparisons, we find that the beam splitting increased the computing time by factors 7.8 and 6.6 although the dosimetric effect was only marginal dose fluctuation from moderate heterogeneity of the round structures.

\begin{table}
\caption{\label{tab:configurations} Definitions of calculation configurations A--F in combinations of with ($+$) or without ($-$) $45^\circ$ tilt $T$, grid normalization $N$, inter-grid sharing $I$, and beam splitting $S$ and total computing times for a broad beam in a heterogeneous phantom.}
\begin{tabular}{cccccc}
\hline\hline
Config. & $T$ & $N$ & $I$ & $S$ & Time (s)\\
\hline
A & $-$ & $-$ & $-$ & $-$ & 1.9\\
B & $-$ & $-$ & $+$ & $-$ & 2.0\\
C & $-$ & $-$ & $+$ & $+$ & 15.5\\
D & $+$ & $-$ & $+$ & $-$ & 3.0\\
E & $+$ & $+$ & $+$ & $-$ & 2.5\\
F & $+$ & $+$ & $+$ & $+$ & 16.4\\
\hline\hline
\end{tabular}
\label{tab:broad}
\end{table}

%

\section{Discussion}

In this work, we introduced grid normalization to successfully resolve the problems with tilted incidence. The interpolation errors are limited by the grid intervals, which may be generally tolerable and controllable because the grid intervals are normally specified by a treatment planner.
We found no apparent loss of speed for interpolation.
In fact, the areal convolution with normal grids was faster than the volumetric convolution with tilted grids.

The beam--grid interplay artifact was overlooked in the original formulation of the GDS algorithm \cite{Kanematsu2008} that was implemented as an extension to the broad-beam algorithm so that the termas and the spreads were calculated exactly at the grids and were free from the interplay.
The inter-grid beam sharing, which is essentially rectangular-kernel convolution, has fully resolved the problem.

As shown in Ref.~\cite{Kanematsu2009b}, the beam splitting resolved the intrinsic problems with the pencil-beam model for fine heterogeneity by dynamic subdivision.
This inevitably requires the interaction-point-of-view approach because otherwise it would be difficult to trace histories of the split beams backwards from each dose-deposition grid \cite{Mackie1985}.
Since the GDS convolution is directly coupled not to individual beams but to resultant terma and spread distributions, beam multiplication due to splitting will not increase the computing time for dose convolution.
In other words, the combination of the GDS and beam-splitting methods is a rational consequence. 

The observed slowing by a factor of several times was mainly attributed to increased transport calculation of the split beams.
These slowing factors were comparable to that originally reported \cite{Kanematsu2009b}, not surprisingly because they in fact used a similar framework of the beam-splitting GDS algorithm though without grid normalization nor inter-grid sharing.
Those examples happened to be of normal incidence and of small interplay and were only intended for a proof of principle of beam splitting.

The terma-weighted mean for the gridded beam spread $\sigma_g$ in Eq.~(\ref{eq:sigmag}) can not be generally valid because the mean spread only approximately represents the contributing Gaussian components.
It was originally assumed that its variation would be small enough to be handled as locally uniform. However, beam splitting changes the spread abruptly.
On the other hand, the beam splitting converts the major part of beam spreads directly into a terma distribution.
As a result, grid-dose spreading handles only the residuals, for which the terma-weighted mean may suffice.

\section{Conclusions}

A known problem of tilted incidence with the original GDS algorithm was naturally resolved by the grid-normalization method without serious loss of accuracy or efficiency.
Another problem of beam--grid interplay artifact was revealed and was resolved by the inter-grid beam-sharing method.

The beam-splitting method for fine-heterogeneity correction will inevitably multiply beams to transport and thus will slow down dose calculation. 
With the GDS algorithm, the dose convolution is made only once after all the beams have been transported, which minimizes the impact of the beam multiplication on computing time.
In fact, for the beams individually split into several tens, the calculation time increased only by several times with the GDS.
This algorithmic framework will thus enable fast and accurate treatment planning of heavy charged particle radiotherapy in the presence of density heterogeneity finer than the size of intrinsic beam blurring.

\end{document}